\documentclass[final,3p,12pt,times]{elsarticle}

\usepackage{amssymb}
\usepackage{amsmath}
\usepackage{ntheorem}
\usepackage{amssymb}
\usepackage{hyperref}
\usepackage{tikz}
\usepackage{xcolor}
\usepackage{hyperref}
\usepackage{graphicx}
\usepackage{subfigure}
\usepackage{graphicx}
\usepackage{setspace}
\usepackage{soul, color, xcolor}    
\usepackage{epstopdf}
\usepackage{epsfig}
\usepackage{balance}
\usepackage{epstopdf}
\usepackage{multicol}
\usepackage{multirow}
\usepackage{threeparttable}
\usepackage{booktabs}
\usepackage{cases}
\usepackage{cuted}
\usepackage{setspace}
\usepackage{makecell} 
\usepackage{algorithm}
\usepackage{algorithmic}

\usepackage{caption}  
\biboptions{sort&compress}

\makeatletter

\newcommand{\Rmnum}[1]{\expandafter\@slowromancap\romannumeral #1@}
\makeatother

\begin{document}
\begin{sloppypar}
	
\begin{frontmatter}



\title{Least-Squares Adaptive Filter-Based Cohen's Class Time-Frequency Distribution for Signal Denoising}

\tnotetext[mytitlenote]{This work was supported in part by the Open Foundation of Hubei Key Laboratory of Applied Mathematics (Hubei University) under Grant HBAM202404; in part by the Foundation of Key Laboratory of System Control and Information Processing, Ministry of Education under Grant Scip20240121; and in part by the Startup Foundation for Introducing Talent of Nanjing Institute of Technology under Grant YKJ202214.}

\author[1]{Manjun Cui}
\author[1,2,3]{Zhichao Zhang\corref{cor1}}\ead{zzc910731@163.com}
\author[4,5]{Yangfan He}
\cortext[cor1]{Corresponding author; Tel: +86-13376073017.}
\address[1]{School of Mathematics and Statistics, Center for Applied Mathematics of Jiangsu Province, and Jiangsu International Joint Laboratory on System Modeling and Data Analysis, Nanjing University of Information Science and Technology, Nanjing 210044, China}
\address[2]{Hubei Key Laboratory of Applied Mathematics, Hubei University, Wuhan 430062, China}
\address[3]{Key Laboratory of System Control and Information Processing, Ministry of Education, Shanghai Jiao Tong University, Shanghai 200240, China}
\address[4]{School of Communication and Artificial Intelligence, School of Integrated Circuits, Nanjing Institute of Technology, Nanjing 211167, China}
\address[5]{Jiangsu Province Engineering Research Center of IntelliSense Technology and System, Nanjing 211167, China}

\begin{abstract}
	Inspired by the use of adaptive kernel-based Cohen's class time-frequency distributions (CCTFDs) for cross-term suppression, this paper aims to explore novel adaptive kernel functions for denoising, {with a particular focus on non-stationary signal processing in practical applications}. We integrate Wiener filter principle and the time-frequency filtering mechanism of CCTFD to design the least-squares adaptive filter method in the  Wigner-Ville distribution (WVD) domain, giving birth to the least-squares adaptive filter-based CCTFD whose kernel function can be adjusted with the input signal automatically to achieve the minimum mean-square error denoising in the WVD domain. {Numerical experiments on typical simulated radar signals  and real-world electrocardiogram data comprehensively demonstrate that the proposed adaptive CCTFD outperforms several state-of-the-art methods in noise suppression.}
\end{abstract}




\end{frontmatter}


\section{Introduction}\label{sec:1}
\indent The field of time-frequency analysis has evolved significantly, offering a diverse array of tools for characterizing non-stationary signals. Traditional linear transforms, such as the short-time Fourier transform (STFT) and the continuous wavelet transform (CWT), provide fundamental frameworks. Over the years, numerous advanced techniques have been developed to enhance resolution and adaptivity, as detailed in the comprehensive text \cite{Coh95}. Among these, Cohen's class time-frequency distribution (CCTFD) \cite{Coh95}, also known as the bi-linear kernel function time-frequency distribution, {stands as one of the most representative tools among conventional time-frequency distributions.} {Notable particular cases include} the Wigner-Ville distribution (WVD)  \cite{wigner1932quantum}, the Choi-Williams distribution \cite{zhao2021research}, the Kirkwood-Rihaczek distribution \cite{rihaczek1968signal}, the Born-Jordan distribution \cite{quan2021novel}, the Zhao-Atlas-Marks distribution \cite{zhao1990use}, the Margenau-Hill distribution \cite{liang2024spatial}, and the Page distribution \cite{xia1996generalized}. Indeed, it can be regarded as a unified bi-linear time-frequency distribution and has found many applications in seismic exploration, electronic countermeasures, deep-sea detection, spectral imaging, and ultrasonic inspection \cite{brynolfsson2021time,wang2021matching,xu2023dtftcnet,vitor2023induction,han2021passive,ding2023regional}.

\indent In addition to traditional CCTFD, recent advancements in time-frequency analysis have focused on sophisticated decomposition techniques and deep learning models. For instance, Bhattacharyya et al. \cite{bhattacharyya2018fourier} proposed an empirical wavelet transform (EWT) based on Fourier-Bessel series expansion (FBSE) to improve boundary detection in the signal spectrum. Similarly, Pachori et al. \cite{pachori2006analysis} combined the FBSE with the WVD to suppress cross-terms {and have also explored} the tunable-Q wavelet transform for cross-term reduction in WVD \cite{pachori2016cross}. In parallel, eigenvalue decomposition methods involving the Hankel matrix have been developed for complex signal analysis \cite{sharma2018eigenvalue}. More recently, an iterative eigenvalue decomposition of the Hankel matrix, functioning as an empirical mode decomposition (EMD)-like tool, was introduced to extract mono-components with high fidelity \cite{singh2025iterative}, and the Fourier-Bessel decomposition method (FBDM) has shown effectiveness in sleep stage classification \cite{gupta2021fbdm}. Meanwhile, deep learning-based adaptive time-frequency analysis methods have attracted attention for their ability to learn optimal representations from data. {These methods have proven highly effective in signal filtering}, denoising, and separation, and have shown effectiveness in tasks such as video, speech, and radar signal analysis \cite{guo2023deep,pan2022tfa,li2023incipient,samavat2022deep,hamdaoui2023improved,zhang2024adaptive}. While these methods represent the state-of-the-art in decomposing signals, they often rely on specific basis functions or iterative sifting processes. In contrast, time-frequency filtering frameworks like CCTFD provide a unified structure that enables direct manipulation of representations, which is particularly suitable for statistically optimal denoising. 

\indent The integral form of CCTFD can be written as the Fourier transform (FT) of the product of the ambiguity function and the kernel function. {Due to the bi-linear nature of CCTFD, cross-terms inevitably arise, causing interference and reducing clarity in signal analysis. Numerous attempts have been made to use the kernel of CCTFD to suppress these cross-terms, yielding various effective kernel functions that significantly enhance CCTFD performance \cite{ning2009repression,martinez2023tunable,qu2020radar,yao2012reducing,lopac2021detection,jadhav2022automated,li2023time}. However, these methods typically employ fixed kernel functions tailored to specific input signals, which severely limits their broader applicability. To overcome this limitation, Baraniuk et al. proposed a series of methods using adaptive kernel functions for cross-term suppression \cite{baraniuk1991radially,212733,258128,jones1995adaptive}, among which the adaptive radial Gaussian kernel is one of the most renowned in this field.} 

{However, in existing literature, adaptive kernel methods have been predominantly designed with the objective of cross-term suppression to improve time-frequency readability, and their direct application explicitly for signal denoising remains insufficiently explored.}{While the low-pass filtering nature inherent in conventional adaptive kernels naturally imparts a certain degree of noise smoothing, their optimization cost functions are predominantly driven by cross-term suppression rather than statistical optimal estimation.These approaches adapt the kernel shape based on the local geometric structure of the ambiguity function to enhance time-frequency concentration and reduce interference between components. Consequently, noise attenuation in these frameworks acts more as an incidental byproduct, since they do not explicitly incorporate noise statistics to guarantee signal reconstruction optimality.} {Because their primary objective is cross-term suppression rather than noise removal, these geometrically motivated kernels inevitably retain noise components that share similar ambiguity domain signatures with the signal auto-terms.}  {As a result, when the signal is heavily contaminated by noise, these geometrically motivated adaptive kernel methods often fail to achieve the desired filtering performance, necessitating a more mathematically rigorous estimation approach.}

{To overcome this limitation, we reformulate the adaptive kernel design problem from the perspective of optimal time-frequency filtering.} The integral form of CCTFD can be rewritten as the conventional convolution of the WVD and the kernel function. It is important to note that the kernel function mentioned here is a Fourier transform pair with the kernel function described above, as shown in Eq.~\eqref{eq:4}. From the {viewpoint} of signal processing, the CCTFD {can be interpreted as a smoothed version of the WVD, achieved via convolution with this kernel.} Inspired by the adaptive kernel function concept proposed by Baraniuk et al., we aim to implement adaptive denoising by designing an adaptive kernel function in the WVD domain. This approach will enable us to effectively extract the target signal from background noise and mitigate the impact of noise through the unique reconstruction property of WVD. 

An adaptive filter \cite{diniz1997adaptive} is a type of filter that automatically adjusts its parameters in response to changes in the signal, making it particularly effective for noise reduction by adapting to varying characteristics of the noise and effectively suppressing it.
The most celebrated result in this field is Wiener's result \cite{wiener1949extrapolation} which applies the minimum mean-square error (MSE) criterion to design the least-squares adaptive filter method. The Wiener filter has been widely used in addressing practical issues {across diverse domains, including} radar, communications, sonar, biomedicine, and vibration engineering \cite{ceccato2020spatial,baudais2023doppler,plabst2020wiener,song2020two,chang2022tensor,qi2021wiener,vered2024parallel}. {Unlike previous adaptive kernel designs focused on cross-term suppression, the proposed method establishes a framework that bridges time-frequency representation and adaptive filtering theory, thereby providing a fundamental advance in both interpretability and performance.}

The core idea of this paper is to {integrate the Wiener filtering principle with} the time-frequency filtering mechanism of CCTFD to investigate the convolution type of CCTFD time-frequency analysis method. The main purpose of this paper is to design the least-squares adaptive filter method in the WVD domain, {elucidate how the kernel function influences denoising performance}, and {derive an optimal adaptive CCTFD kernel that achieves the theoretical minimum MSE}. The main contributions of this paper are summarized as follows:
\begin{itemize}
	\item This paper obtains the least-squares adaptive filter in the WVD domain.
	
	\item {This paper proposes an adaptive CCTFD, whose kernel function adopts the reversal of the least-squares adaptive filter transfer function in the WVD domain, to effectively restore signals from severe additive noise backgrounds (with the signal-to-noise ratio (SNR) ranging from $-10$ dB to $5$ dB).}
	
	\item This paper demonstrates the noise suppression superiority of the proposed adaptive CCTFD over the ordinary Wiener filter, some classical fixed kernel function-based CCTFDs, and the adaptive radial Gaussian kernel function-based CCTFD. Furthermore, it outperforms representative neural network-based approaches on real-world data.
\end{itemize}

The remainder of this paper is structured as follows. In Section~\ref{sec:2}, we {review} some necessary background and notation on the convolution type of CCTFD. In Section~\ref{sec:3}, we provide the CCTFD-based {adaptive filtering method for signals corrupted by additive noise}. In Section~\ref{sec:4}, we introduce {numerical examples and real-world data experiments} to validate the effectiveness, reliability, and feasibility of the proposed method. In Section~\ref{sec:5}, we draw a conclusion. {Finally, Section~\ref{sec:6} discusses potential extensions and future research directions of the proposed framework. }All the technical proofs of our theoretical results are relegated to the appendix parts.

\section{Convolution type of CCTFD}\label{sec:2}
{ To facilitate the understanding and clear representation of the various symbols used throughout the paper, we present Table~1, which provides a detailed explanation of each symbol and its corresponding meaning.}

\begin{table} 
	\centering
	\caption{ Symbol description.}
		\begin{tabular}{ll}\hline
			{Symbol} & Description  \\ \hline
			$\mathrm{T}$ & transpose operator  \\
			--- & complex conjugate operator \\
			$\ast$ & convolution operator \\
			$\Vert\cdot\Vert_2$ & $L^2$-norm operator \\
			$\mathbb{E}(\cdot)$ & mathematical expectation operator \\
			$R$ & correlation function operator \\
			$\mathrm{C}$ & CCTFD operator \\
			$\mathrm{W}$ & WVD operator \\
			$\mathcal{F}$ & FT operator\\
			$\phi$ & kernel function \\
			$\Pi$ & FT of the kernel function $\phi$ \\
			$H$  & adaptive filter in the WVD domain \\
			$\sigma_{\mathrm{MSE}}^2$ & MSE \\
			$\varepsilon$ & PSD \\
			$\mathrm{C}_g^{\mathrm{LSAF}}$ & {the least-squares adaptive filter-based CCTFD for the input signal $g$} \\
			\hline
		\end{tabular}
\end{table}

\indent The integral form of CCTFD of the function $f({\boldsymbol{t}})\in L^2(\mathbb{R}^{N})$ reads \cite{Coh95,ozaktas1996effect}
\begin{align}\label{eq:1}
	\mathrm{C}_f({\boldsymbol{t}},\boldsymbol{w}) 
	=&\int_{\mathbb{R}^N}\int_{\mathbb{R}^N}\int_{\mathbb{R}^N}f\left(\boldsymbol{y}+\frac{\boldsymbol{\tau}}{2}\right)\overline{f\left(\boldsymbol{y}-\frac{\boldsymbol{\tau}}{2}\right)} \phi(\boldsymbol{\theta},\boldsymbol{\tau})\mathrm{e}^{-2\pi\mathrm{i}\left(\boldsymbol{\theta}{\boldsymbol{t}}^\mathrm{T}+\boldsymbol{\tau}\boldsymbol{w}^\mathrm{T}-\boldsymbol{y}\boldsymbol{\theta}^\mathrm{T}\right)}\mathrm{d}\boldsymbol{y}\mathrm{d}\boldsymbol{\tau}\mathrm{d}\boldsymbol{\theta},
\end{align}
where the superscripts $\mathrm{T}$ and --- denote the transpose operator and complex conjugate operator, respectively, and $\phi(\boldsymbol{\theta},\boldsymbol{\tau})$ denotes the kernel function.

Let $\ast$ and $\mathcal{F}$ be the conventional convolution operator and Fourier operator, respectively. Then, Eq.~\eqref{eq:1} can be rewritten as
\begin{align}\label{eq:2}
	\mathrm{C}_f({\boldsymbol{t}},\boldsymbol{w})=&\left(\mathrm{W}_{f}\ast\Pi\right)({\boldsymbol{t}},\boldsymbol{w}),
\end{align}
where
\begin{equation}\label{eq:3}
	\mathrm{W}_f({\boldsymbol{t}},\boldsymbol{w})=\int_{\mathbb{R}^N}f\left({\boldsymbol{t}}+\frac{\boldsymbol{\tau}}{2}\right)\overline{f\left({\boldsymbol{t}}-\frac{\boldsymbol{\tau}}{2}\right)}\mathrm{e}^{-2\pi\mathrm{i}\boldsymbol{\tau}\boldsymbol{w}^\mathrm{T}}\mathrm{d}\boldsymbol{\tau}
\end{equation}
denotes the WVD of the function $f({\boldsymbol{t}})$, and
\begin{align}\label{eq:4}
	\Pi({\boldsymbol{t}},\boldsymbol{w})
	=\mathcal{F}[\phi]({\boldsymbol{t}},\boldsymbol{w}) 
	=\int_{\mathbb{R}^N}\int_{\mathbb{R}^N}\phi(\boldsymbol{\theta},\boldsymbol{\tau})\mathrm{e}^{-2\pi\mathrm{i}\left(\boldsymbol{\theta}{\boldsymbol{t}}^\mathrm{T}+\boldsymbol{\tau}\boldsymbol{w}^\mathrm{T}\right)}\mathrm{d}\boldsymbol{\theta}\mathrm{d}\boldsymbol{\tau}
\end{align}
denotes the FT of the kernel function $\phi(\boldsymbol{\theta},\boldsymbol{\tau})$. Eq.~\eqref{eq:2} indicates that the CCTFD is {equivalent to} the convolution of the WVD and the (FT version of) kernel function.

{ To provide a concrete illustration, some typical kernel functions $\phi(\boldsymbol{\theta},\boldsymbol{\tau})$ corresponding to well-known Cohen's class distributions are summarized in Table \ref{tab:kernel_functions}. By selecting different kernel functions, the CCTFD can be tailored to specific signal analysis requirements.}

\begin{table}[htbp]
	
	\centering
	\caption{\label{tab:kernel_functions}Typical kernel functions for various CCTFDs.}
	\footnotesize
	\begin{tabular}{cc}
		\specialrule{0.1em}{4pt}{4pt}
		$\phi(\boldsymbol{\theta},\boldsymbol{\tau})$ & CCTFD\\
		\specialrule{0.1em}{4pt}{4pt}
		$1$ & WD\\
		\specialrule{0em}{4pt}{4pt}
		$\mathrm{e}^{-\frac{\lVert\boldsymbol{\theta}\rVert^2\lVert\boldsymbol{\tau}\rVert^2}{\varsigma}}$ & Choi-Williams distribution\\
		\specialrule{0em}{4pt}{4pt}
		$\mathrm{e}^{\mathrm{i}\frac{\boldsymbol{\theta}\boldsymbol{\tau}^\mathrm{T}}{2}}$ & Kirkwood-Rihaczek distribution\\
		\specialrule{0em}{4pt}{4pt}
		$\frac{\sin\left(\frac{\boldsymbol{\theta}\boldsymbol{\tau}^\mathrm{T}}{2}\right)}{\frac{\boldsymbol{\theta}\boldsymbol{\tau}^\mathrm{T}}{2}}$ & Born-Jordan distribution\\
		\specialrule{0em}{4pt}{4pt}
		$g(\boldsymbol{\tau})\left\|\boldsymbol{\tau}\right\|_1\frac{\sin\left(2\pi\kappa\boldsymbol{\theta}\boldsymbol{\tau}^\mathrm{T}\right)}{2\pi\kappa\boldsymbol{\theta}\boldsymbol{\tau}^\mathrm{T}}$ & Zhao-Atlas-Marks distribution\\
		\specialrule{0em}{4pt}{4pt}
		$\cos\left(\frac{\boldsymbol{\theta}\boldsymbol{\tau}^\mathrm{T}}{2}\right)$ & Margenau-Hill distribution\\
		\specialrule{0em}{4pt}{4pt}
		$\mathrm{e}^{\mathrm{i}\boldsymbol{\theta}\left\|\boldsymbol{\tau}\right\|_1}$ & Page distribution\\
		\specialrule{0.1em}{4pt}{4pt}
	\end{tabular}
\end{table}

\section{Least-squares adaptive filter-based CCTFD}\label{sec:3}
\indent For a given noise polluted signal $g({\boldsymbol{t}})=f({\boldsymbol{t}})+n({\boldsymbol{t}})$, where $f({\boldsymbol{t}})$ and $n({\boldsymbol{t}})$ denote the pure signal and the additive noise, respectively. The common tactic of filters is to restore the pure signal as accurately as possible, namely, {to find an estimate $\hat{f}(x)$ that is} as close as possible to the ideal $f({\boldsymbol{t}})$. {Owing to} the convolution nature of CCTFD and the unique reconstruction property of WVD, this is equivalent to {designing} an adaptive filter $H({\boldsymbol{t}},\boldsymbol{w})$ in the WVD domain, which can find the estimate
\begin{equation}\label{eq:5}
	\mathrm{W}_{\widehat{f}}({\boldsymbol{t}},\boldsymbol{w})=\left(\mathrm{W}_g\ast H\right)({\boldsymbol{t}},\boldsymbol{w})
\end{equation}
as close as possible to the ideal $\mathrm{W}_f({\boldsymbol{t}},\boldsymbol{w})$. According to Wiener filter principle, a natural criterion to characterize the estimation accuracy is the MSE criterion
\begin{equation}\label{eq:6}
	\sigma_{\mathrm{MSE}}^2\overset{\mathrm{def}}{=}\mathbb{E}\left\{\left|\mathrm{W}_f({\boldsymbol{t}},\boldsymbol{w})-\mathrm{W}_{\widehat{f}}({\boldsymbol{t}},\boldsymbol{w})\right|^2\right\},
\end{equation}
where $\mathbb{E}(\cdot)$ denotes the mathematical expectation operator.

For simplicity, let $\boldsymbol{z}=({\boldsymbol{t}},\boldsymbol{w})\in\mathbb{R}^{2N}$, then Eqs.~\eqref{eq:5} and \eqref{eq:6} become
\begin{align}\label{eq:7}
	\mathrm{W}_{\widehat{f}}(\boldsymbol{z})=\left(\mathrm{W}_g\ast H\right)(\boldsymbol{z})
	=\int_{\mathbb{R}^{2N}}\mathrm{W}_g(\boldsymbol{k})H(\boldsymbol{z}-\boldsymbol{k})\mathrm{d}\boldsymbol{k}
\end{align}
and
\begin{equation}\label{eq:8}
	\sigma_{\mathrm{MSE}}^2=\mathbb{E}\left\{\left|\mathrm{W}_f(\boldsymbol{z})-\mathrm{W}_{\widehat{f}}(\boldsymbol{z})\right|^2\right\},
\end{equation}
respectively. Now, our goal is to design an adaptive optimal filter $H_{\mathrm{opt}}(\boldsymbol{z})$ in the WVD domain to minimize the MSE given by Eq.~\eqref{eq:8}, or equivalently,
\begin{equation}\label{eq:9}
	H_{\mathrm{opt}}(\boldsymbol{z})=\mathop{\arg\min}\limits_{H(\boldsymbol{z})}\sigma_{\mathrm{MSE}}^{2}.
\end{equation}
\indent By using the orthogonal principle \cite{proakis1985probability}, the stationary assumption and the conventional convolution and correlation theorems to establish, simplify and solve the Wiener-Hopf equation, respectively, the least-squares adaptive filter transfer function in the WVD domain reads
\begin{equation}\label{eq:10}
	\mathcal{F}\left[H_{\mathrm{opt}}\right](\boldsymbol{u})=\frac{\varepsilon_{\mathrm{W}_f,\mathrm{W}_g}(\boldsymbol{u})}{\varepsilon_{\mathrm{W}_g}(\boldsymbol{u})},
\end{equation}
where $\varepsilon_{\mathrm{W}_f,\mathrm{W}_g}(\boldsymbol{u})=\mathcal{F}\left[\mathrm{W}_f\right](\boldsymbol{u})\overline{\mathcal{F}\left[\mathrm{W}_g\right](\boldsymbol{u})}$ denotes the power spectral density (PSD) of $\mathrm{W}_f(\boldsymbol{z})$ and $\mathrm{W}_g(\boldsymbol{z})$, and $\varepsilon_{\mathrm{W}_g}(\boldsymbol{u})=\left|\mathcal{F}\left[\mathrm{W}_g\right](\boldsymbol{u})\right|^{2}$ denotes the PSD of $\mathrm{W}_g(\boldsymbol{z})$. 

{To theoretically justify this derivation, the additive noise is explicitly assumed to be wide-sense stationary (WSS). Furthermore, while the target signals are inherently non-stationary, their distributions $\mathrm{W}_f$ and $\mathrm{W}_g$ are treated as macroscopically stationary random fields over the observation block. As detailed in Appendix A, this standard theoretical idealization enables the analytical, closed-form solution of the 2D Wiener-Hopf equation.} Taking the inverse FT on both sides of Eq.~\eqref{eq:10} yields the least-squares adaptive filter in the WVD domain
\begin{equation}\label{eq:11}
	H_{\mathrm{opt}}(\boldsymbol{z})=\int_{\mathbb{R}^{2N}}\frac{\varepsilon_{\mathrm{W}_f,\mathrm{W}_g}(\boldsymbol{u})}{\varepsilon_{\mathrm{W}_g}(\boldsymbol{u})}\mathrm{e}^{2\pi\mathrm{i}\boldsymbol{u}\boldsymbol{z}^\mathrm{T}}\mathrm{d}\boldsymbol{u}.
\end{equation}

See Appendix~A for the detailed derivation of Eq.~\eqref{eq:10}. Correspondingly, the minimum MSE can be reduced to zero, i.e.
\begin{equation}\label{eq:12}
	\mathop{\min}\limits_{H(\boldsymbol{z})}\sigma_{\mathrm{MSE}}^{2}=0,
\end{equation}

See Appendix~B for the detailed derivation of Eq.~\eqref{eq:12}.

From Eq.~\eqref{eq:4}, it derives that the adaptive optimal kernel function takes the reversal of the least-squares adaptive filter transfer function in the WVD domain, i.e.,
\begin{equation}\label{eq:13}
	\phi_{\mathrm{opt}}(\boldsymbol{\theta},\boldsymbol{\tau})=\mathcal{F}\left[H_{\mathrm{opt}}\right](-\boldsymbol{\theta},-\boldsymbol{\tau})=\frac{\varepsilon_{\mathrm{W}_f,\mathrm{W}_g}(-\boldsymbol{\theta},-\boldsymbol{\tau})}{\varepsilon_{\mathrm{W}_g}(-\boldsymbol{\theta},-\boldsymbol{\tau})}.
\end{equation}
\indent Substituting Eq.~\eqref{eq:13} into Eq.~\eqref{eq:1} gives the least-squares adaptive filter-based CCTFD
\begin{align}\label{eq:14}
	\mathrm{C}_g^{\mathrm{LSAF}}({\boldsymbol{t}},\boldsymbol{w}) 
	=\int_{\mathbb{R}^N}\int_{\mathbb{R}^N}\int_{\mathbb{R}^N}g\left(\boldsymbol{y}+\frac{\boldsymbol{\tau}}{2}\right)\overline{g\left(\boldsymbol{y}-\frac{\boldsymbol{\tau}}{2}\right)} \frac{\varepsilon_{\mathrm{W}_f,\mathrm{W}_g}(-\boldsymbol{\theta},-\boldsymbol{\tau})}{\varepsilon_{\mathrm{W}_g}(-\boldsymbol{\theta},-\boldsymbol{\tau})}\mathrm{e}^{-2\pi\mathrm{i}\left(\boldsymbol{\theta}{\boldsymbol{t}}^\mathrm{T}+\boldsymbol{\tau}\boldsymbol{w}^\mathrm{T}-\boldsymbol{y}\boldsymbol{\theta}^\mathrm{T}\right)}\mathrm{d}\boldsymbol{y}\mathrm{d}\boldsymbol{\tau}\mathrm{d}\boldsymbol{\theta}.
\end{align}

It is obvious that the transformation is designed to adapt based on the input signals $f$ and $g$. { This adaptability is achieved by analytically deriving the optimal CCTFD kernel function directly from the cross-power spectral density of the input signals in the WVD domain. Consequently, the filter dynamically shapes itself to varying signal conditions without the need for heuristic iterative algorithms, providing a statistically optimal filtering process that ultimately enhances the clarity and accuracy of the time-frequency representation. To explicitly illustrate the technical details of this strategy and enhance reproducibility, the step-by-step implementation of the proposed adaptive CCTFD is summarized in Algorithm \ref{alg:LSAF_CCTFD}.}

\begin{algorithm}[htbp]
	
	\caption{Least-squares adaptive filter-based CCTFD generation}
	\label{alg:LSAF_CCTFD}
	\begin{algorithmic}[1]
		\REQUIRE Noisy observation $g(\boldsymbol{t})$, pure signal $f(\boldsymbol{t})$.
		\ENSURE Denoised time-frequency representation $C_g^{\mathrm{LSAF}}(\boldsymbol{t},\boldsymbol{w})$.
		
		\STATE \textbf{Phase 1: Time-frequency transformation}
		\STATE Compute the WVD of  $f$ via Eq.~\eqref{eq:3}: $W_f(\boldsymbol{t},\boldsymbol{w})$.
		\STATE Compute the WVD of $g$ via Eq.~\eqref{eq:3}: $W_g(\boldsymbol{t},\boldsymbol{w})$ .
		
		\STATE \textbf{Phase 2: Statistical estimation}
		\STATE Estimate the cross-PSD of $W_f$ and $W_g$: $\varepsilon_{\mathrm{W}_f,\mathrm{W}_g}(\boldsymbol{u}) = \mathcal{F}[\mathrm{W}_f](\boldsymbol{u})\overline{\mathcal{F}[\mathrm{W}_g](\boldsymbol{u})}$.
		\STATE Estimate the self-PSD of $W_g$: $\varepsilon_{\mathrm{W}_g}(\boldsymbol{u}) = \left| \mathcal{F}[\mathrm{W}_g](\boldsymbol{u}) \right|^2$.
		
		\STATE \textbf{Phase 3: Filter derivation \& Kernel synthesis}
		\STATE Construct the least-squares adaptive filter transfer function $\mathcal{F}[H_{opt}]$ according to Eq.~\eqref{eq:10}.
		\STATE \emph{Objective: Achieve the optimal theoretical minimum MSE ($\min\sigma_{MSE}^2=0$) as per Eq.~\eqref{eq:12}.}
		\STATE Obtain the adaptive filter $H_{opt}$ in the WVD domain via inverse FT based on Eq.~\eqref{eq:11}.
		\STATE Synthesize the optimal adaptive kernel function $\phi_{opt}$ based on Eq.~\eqref{eq:13}.
		
		\STATE \textbf{Phase 4: Optimal reconstruction}
		\STATE Apply the optimal kernel $\phi_{opt}$ to convolve with the WVD of the input signal.
		\STATE Compute the final denoised distribution $C_g^{\mathrm{LSAF}}(\boldsymbol{t},\boldsymbol{w})$ via Eq.~\eqref{eq:14}.
		
		\RETURN $C_g^{\mathrm{LSAF}}(\boldsymbol{t},\boldsymbol{w})$
	\end{algorithmic}
\end{algorithm}

The proposed least-squares adaptive CCTFD differs conceptually from several recent time-frequency analysis advancements. While methods based on FBSE \cite{bhattacharyya2018fourier, pachori2006analysis, gupta2021fbdm} and tunable-Q wavelet transform \cite{pachori2016cross} focus on decomposing multi-component signals to mitigate interference, our approach targets optimal denoising via the minimum MSE criterion within the CCTFD framework. Unlike Hankel matrix-based techniques \cite{sharma2018eigenvalue, singh2025iterative}, which rely on iterative eigenvalue decompositions or sifting processes to extract components, the proposed method integrates Wiener filtering theory directly into the time-frequency domain. {It should be noted that while practical adaptive filters often employ iterative convergence algorithms (e.g., Least Mean Squares) to approximate the optimal weights, the classical Wiener theory yields a closed-form analytical solution based on power spectral densities. By analytically synthesizing this optimal kernel, this establishes a direct, non-iterative, filtering-based paradigm that emphasizes statistical optimality under the minimum MSE criterion.} From this perspective, the proposed method is complementary to decomposition-oriented tools: while FBSE-, tunable-Q-, and Hankel matrix-based approaches aim to extract intrinsic signal components, our approach focuses on optimal denoising and reconstruction directly in the time-frequency domain.

\section{Numerical experiments}\label{sec:4}
\indent {In this section, four numerical examples are presented} to verify the correctness and effectiveness of the least-squares adaptive filter method in the WVD domain. {The experiments include both simulated signals and real-world data from the {electrocardiogram (ECG) dataset \cite{ECG_dataset}}. The denoising performance of the proposed adaptive CCTFD is compared with that of} the ordinary Wiener filter, some classical fixed kernel function-based CCTFDs and the adaptive radial Gaussian kernel function-based CCTFD, while representative neural network-based methods are considered for comparison on real-world radar data.

In simulations, the fixed kernel functions are chosen as $\phi(\boldsymbol{\theta},\boldsymbol{\tau})=\cos\left(\frac{\boldsymbol{\theta}\boldsymbol{\tau}^\mathrm{T}}{2}\right)$, $\phi(\boldsymbol{\theta},\boldsymbol{\tau})=\mathrm{e}^{\mathrm{i}\frac{\boldsymbol{\theta}\boldsymbol{\tau}^\mathrm{T}}{2}}$, $\phi(\boldsymbol{\theta},\boldsymbol{\tau})=\frac{\sin\left(\frac{\boldsymbol{\theta}\boldsymbol{\tau}^\mathrm{T}}{2}\right)}{\frac{\boldsymbol{\theta}\boldsymbol{\tau}^\mathrm{T}}{2}}$ and $\phi(\boldsymbol{\theta},\boldsymbol{\tau})=\mathrm{e}^{\mathrm{i}\boldsymbol{\theta}\left\|\boldsymbol{\tau}\right\|_1}$, (here $\left\|\cdot\right\|_1$ denotes the $1$-norm for vectors), corresponding to the Margenau-Hill distribution, the Kirkwood-Rihaczek distribution, the Born-Jordan distribution and the Page distribution, respectively. For simplicity, we {refer to} the filtering methods using fixed kernel functions $\cos\left(\frac{\boldsymbol{\theta}\boldsymbol{\tau}^\mathrm{T}}{2}\right)$, $\mathrm{e}^{\mathrm{i}\frac{\boldsymbol{\theta}\boldsymbol{\tau}^\mathrm{T}}{2}}$, $\frac{\sin\left(\frac{\boldsymbol{\theta}\boldsymbol{\tau}^\mathrm{T}}{2}\right)}{\frac{\boldsymbol{\theta}\boldsymbol{\tau}^\mathrm{T}}{2}}$ and $\mathrm{e}^{\mathrm{i}\boldsymbol{\theta}\left\|\boldsymbol{\tau}\right\|_1}$ as Margenau-Hill, Kirkwood-Rihaczek, Born-Jordan and Page, respectively. Additionally, the filtering method utilizing the adaptive radial Gaussian kernel function-based CCTFD is referred to as the adaptive radial Gaussian kernel.

{
	To quantitatively evaluate the denoising performance, the SNR, MSE, and peak SNR (PSNR) are utilized as the evaluation metrics. Let $T$ be the length of the time series. These metrics are mathematically defined as
	\begin{equation}\label{eq:snr}
		\text{SNR} = 10 \log_{10} \left( \frac{\Vert f \Vert_2^2}{\Vert n \Vert_2^2} \right),
	\end{equation}
	\begin{equation}\label{eq:mse}
		\text{MSE} = \frac{1}{T} \Vert f-\widehat{f} \Vert_2^2,
	\end{equation}
	\begin{equation}\label{eq:psnr}
		\text{PSNR} = 10 \log_{10} \left( \frac{\left( \max|f| \right)^2}{\text{MSE}} \right).
	\end{equation}
}

\emph{Example~1 (Linear frequency-modulated (LFM) signal):} The polluted signal is selected for the LFM signal $\mathrm{e}^{2\pi\mathrm{i}\left({\boldsymbol{t}}+\frac{{\boldsymbol{t}}^{2}}{2}\right)}$.

\emph{Example~2 (Gaussian enveloped LFM (GELFM) signal):} The polluted signal is selected for the GELFM signal $\mathrm{e}^{-\frac{({\boldsymbol{t}}+1)^2}{8}}\mathrm{e}^{2\pi\mathrm{i}{\boldsymbol{t}}^{2}}$.

\emph{Example~3 (Quadratic frequency-modulated (QFM) signal):} The polluted signal is selected for the QFM signal $\mathrm{e}^{2\pi\mathrm{i}\left(-3{\boldsymbol{t}}+\frac{{\boldsymbol{t}}^2}{2}+\frac{{\boldsymbol{t}}^3}{4}\right)}$.

{\emph{Example~4 (Two-component LFM (TC-LFM) signal):} The polluted signal is selected for the TC-LFM signal $\mathrm{e}^{2\pi\mathrm{i}(\boldsymbol{t}+0.5\boldsymbol{t}^{2})} + \mathrm{e}^{2\pi\mathrm{i}(-\boldsymbol{t}+0.5\boldsymbol{t}^{2})}$.}

{In examples 1--4, additive white Gaussian noise is introduced to the signals, with the SNR ranging from $-10\mathrm{dB}$ to $5\mathrm{dB}$ and the observation interval is set to  $[-5\mathrm{s},5\mathrm{s}]$.} { The sampling frequencies are set as follows: $30\mathrm{Hz}$ for example 1, $50\mathrm{Hz}$ for example 2, $150\mathrm{Hz}$ for example 3 and $10\mathrm{Hz}$ for example 4.}

Figures~1(a) and (b) plot respectively the SNR-MSE (logarithm base $10$) and PSNR (average of real and imaginary parts) line charts of the estimated LFM {signal} using seven filtering methods including Margenau-Hill, Kirkwood-Rihaczek, Born-Jordan, Page, adaptive radial Gaussian kernel,  Wiener filter and the proposed adaptive CCTFD. Figures~2(a) and (b) plot respectively the SNR-MSE and SNR-PSNR line charts of the estimate GELFM signals using these seven filtering methods. Figures~3(a) and (b) plot respectively the SNR-MSE and SNR-PSNR line charts of the estimate QFM signals using these seven filtering methods. Figures~4(a) and (b) plot respectively the SNR-MSE and SNR-PSNR line charts of the estimate {TC-LFM} signals using these seven filtering methods. It can be seen that the proposed adaptive CCTFD achieves better noise suppression performance than Margenau-Hill, Kirkwood-Rihaczek, Born-Jordan, Page, adaptive radial Gaussian kernel, and Wiener filters under different SNR levels.

\begin{figure}[!ht]
	\centering
	\subfigure[]{\includegraphics[width=0.9\textwidth]{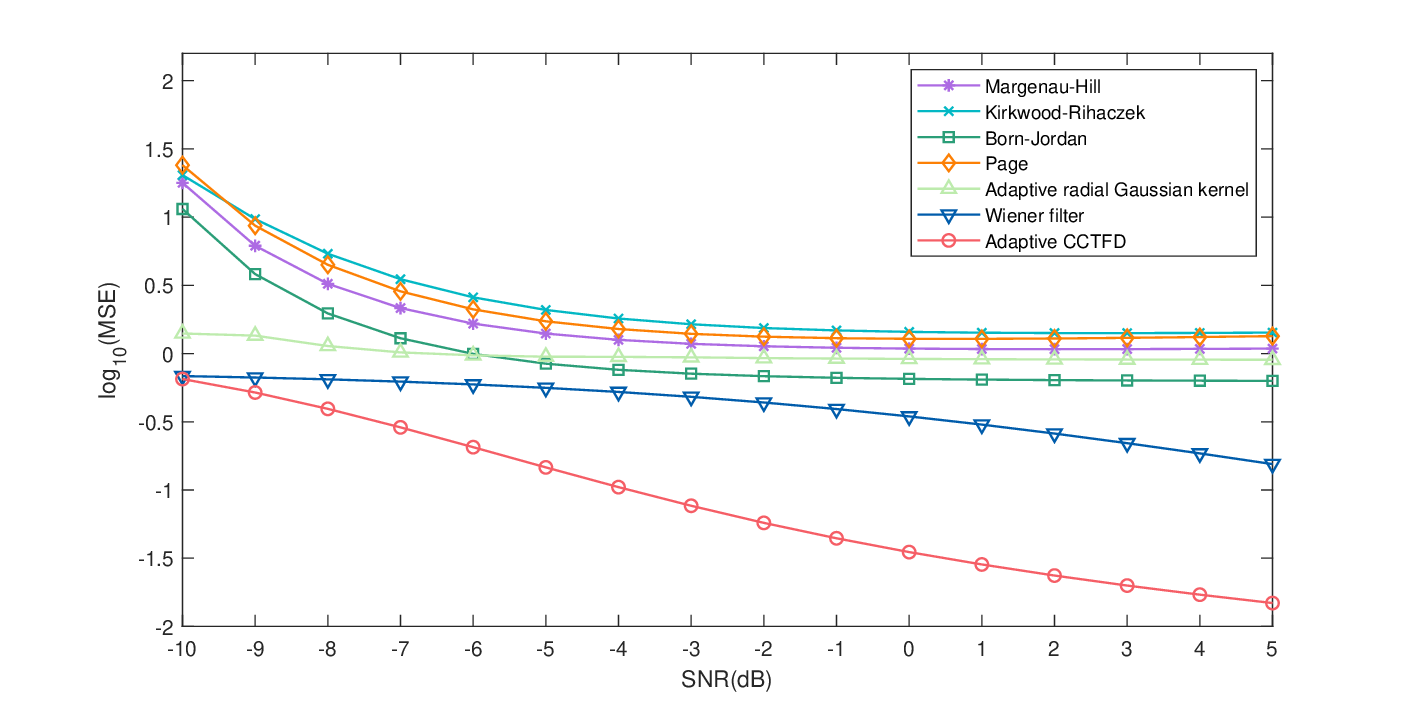}}
	\subfigure[]{\includegraphics[width=0.9\textwidth]{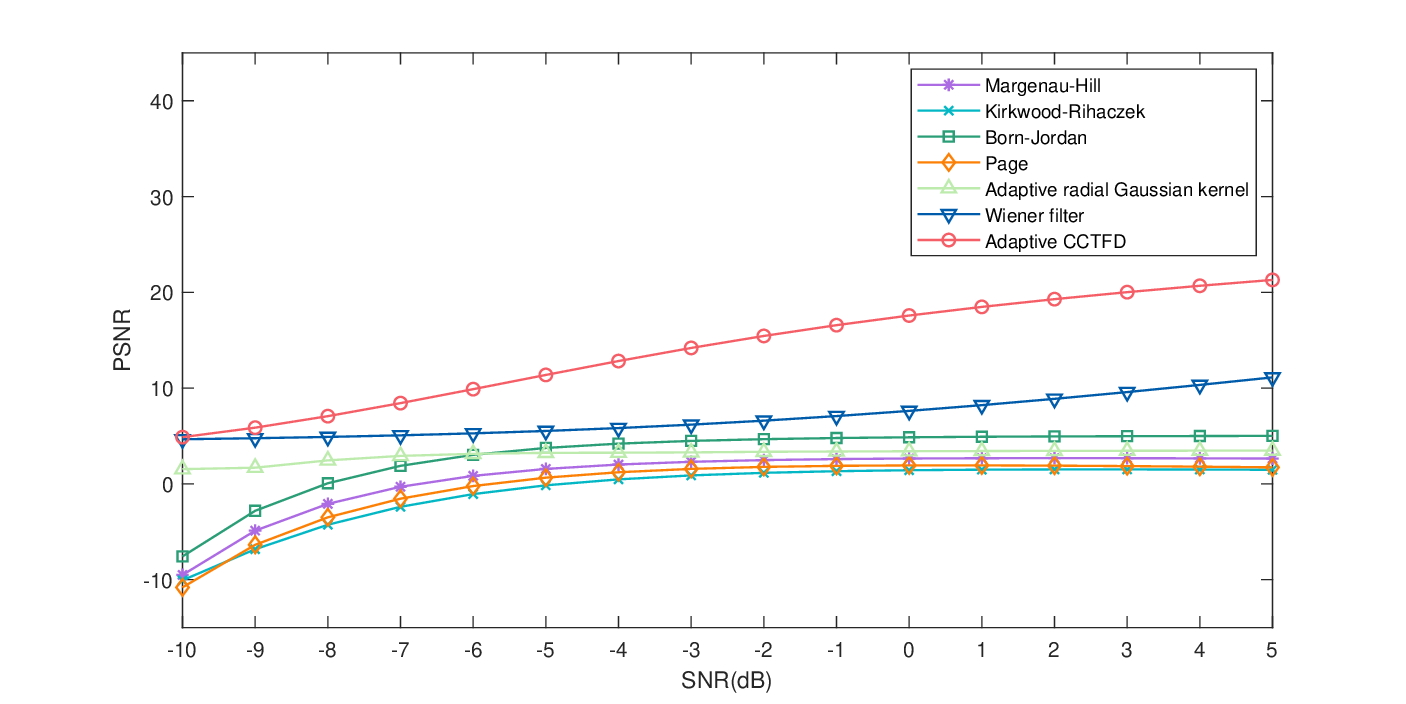}}
	\caption{Performance evaluation of seven denoising methods for the LFM signal at SNR levels from $-10\mathrm{dB}$ to $5\mathrm{dB}$ under white Gaussian noise: (a) $\log_{10}(\text{MSE})$; (b) PSNR.}
	\label{fig1}
\end{figure}

\begin{figure}[!ht]
	\centering
	\subfigure[]{\includegraphics[width=0.9\textwidth]{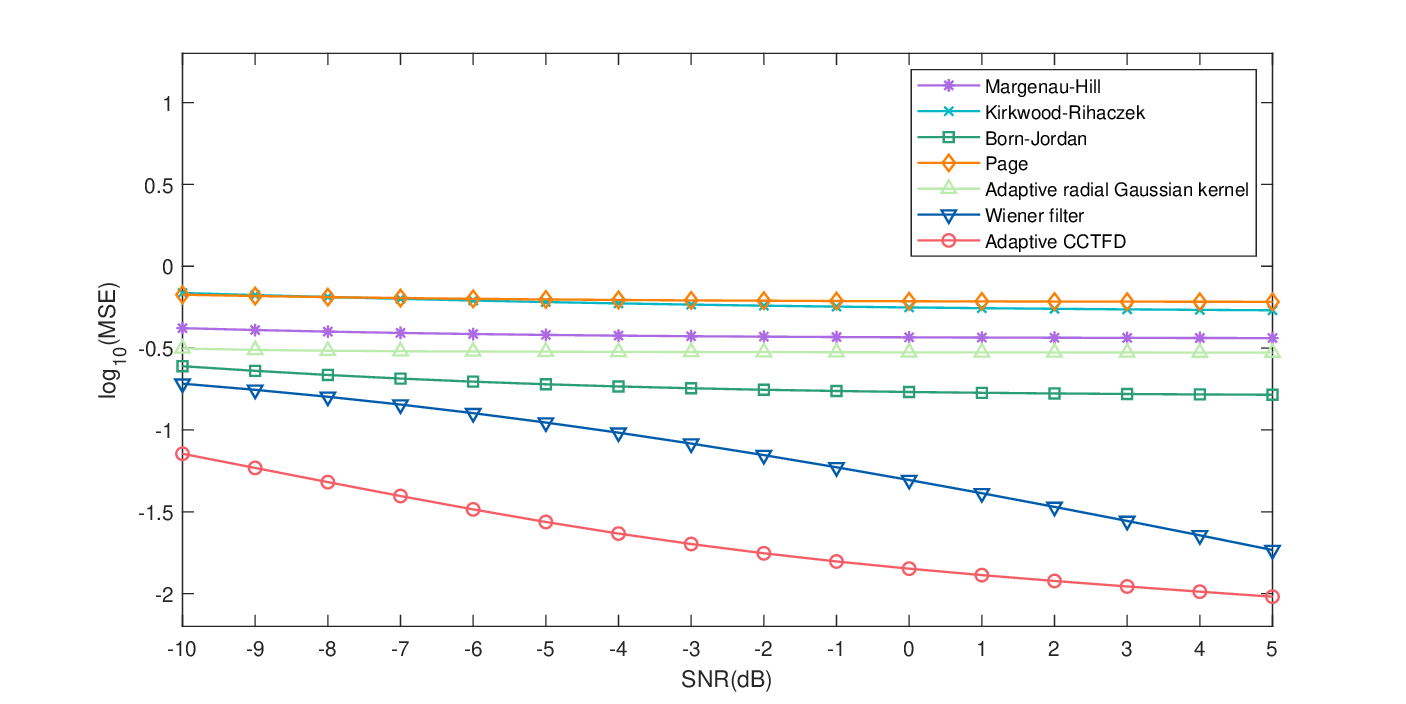}}
	\subfigure[]{\includegraphics[width=0.9\textwidth]{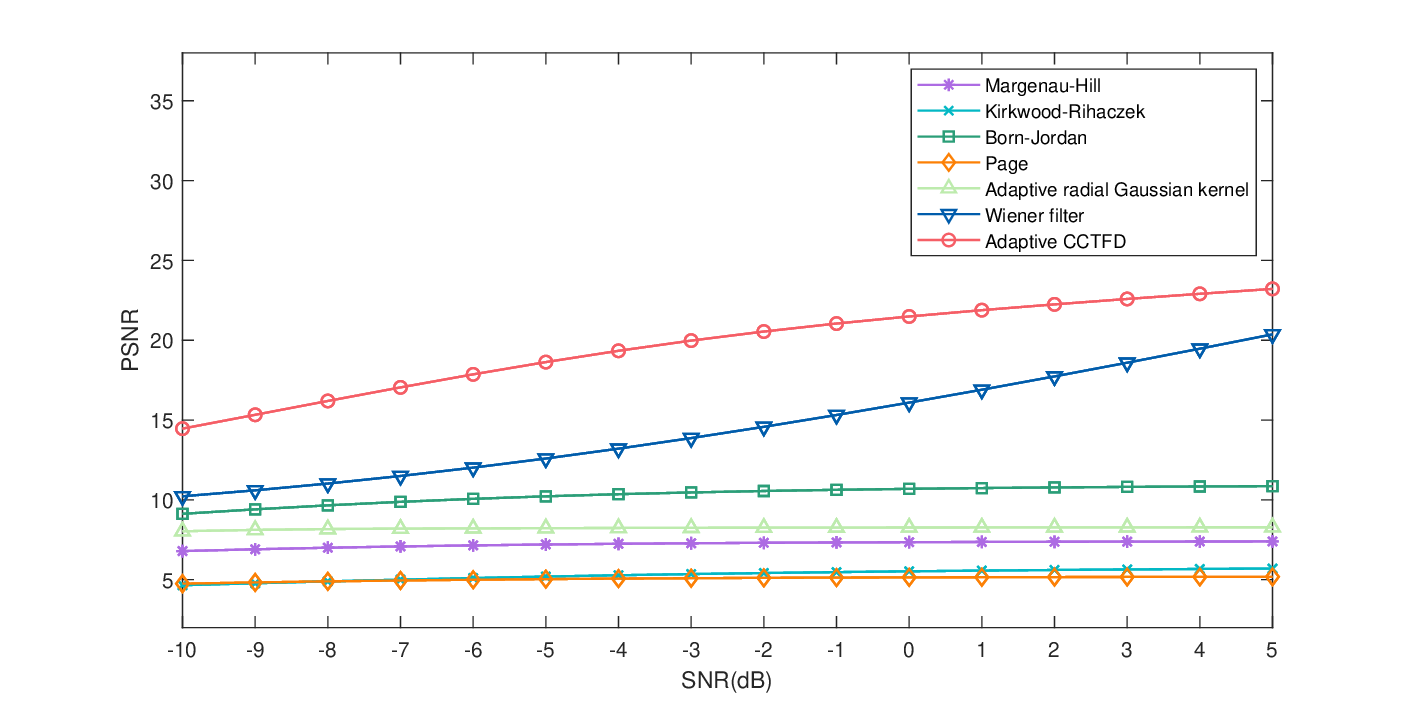}}
	\caption{Performance evaluation of seven denoising methods for the GELFM signal at SNR levels from $-10\mathrm{dB}$ to $5\mathrm{dB}$ under white Gaussian noise: (a) $\log_{10}(\text{MSE})$; (b) PSNR.}
	\label{fig2}
\end{figure}

\begin{figure}[!ht]
	\centering
	\subfigure[]{\includegraphics[width=0.9\textwidth]{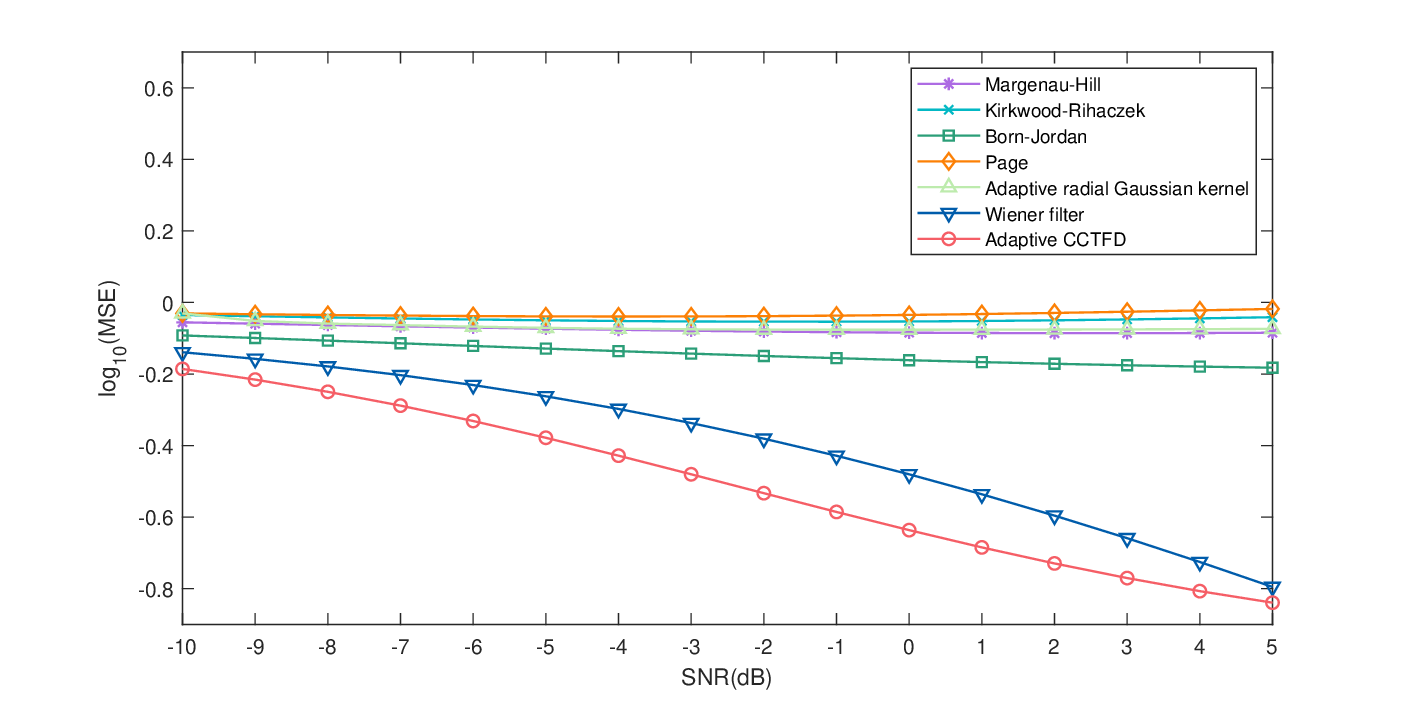}}
	\subfigure[]{\includegraphics[width=0.9\textwidth]{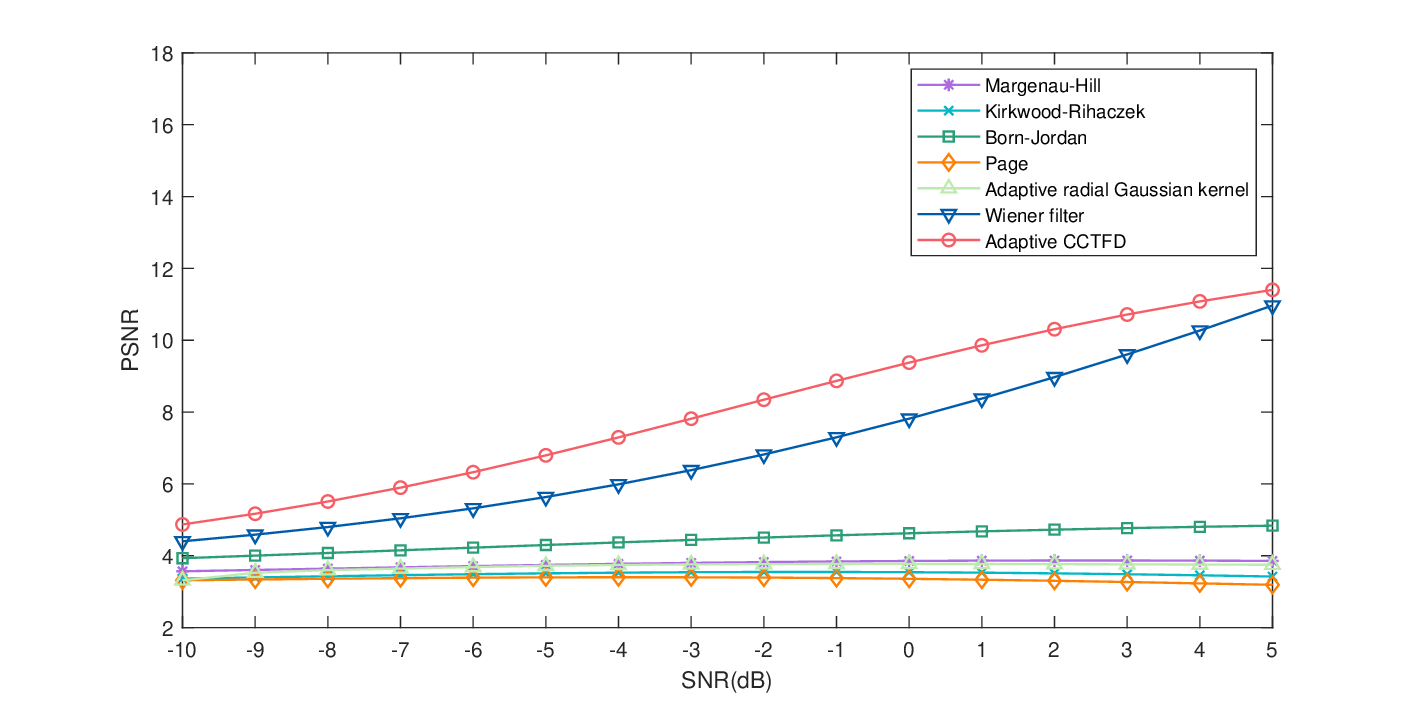}}
	\caption{Performance evaluation of seven denoising methods for the QFM signal at SNR levels from $-10\mathrm{dB}$ to $5\mathrm{dB}$ under white Gaussian noise: (a) $\log_{10}(\text{MSE})$; (b) PSNR.}
	\label{fig3}
\end{figure}

\begin{figure}[!ht]
	\centering
	\subfigure[]{\includegraphics[width=0.9\textwidth]{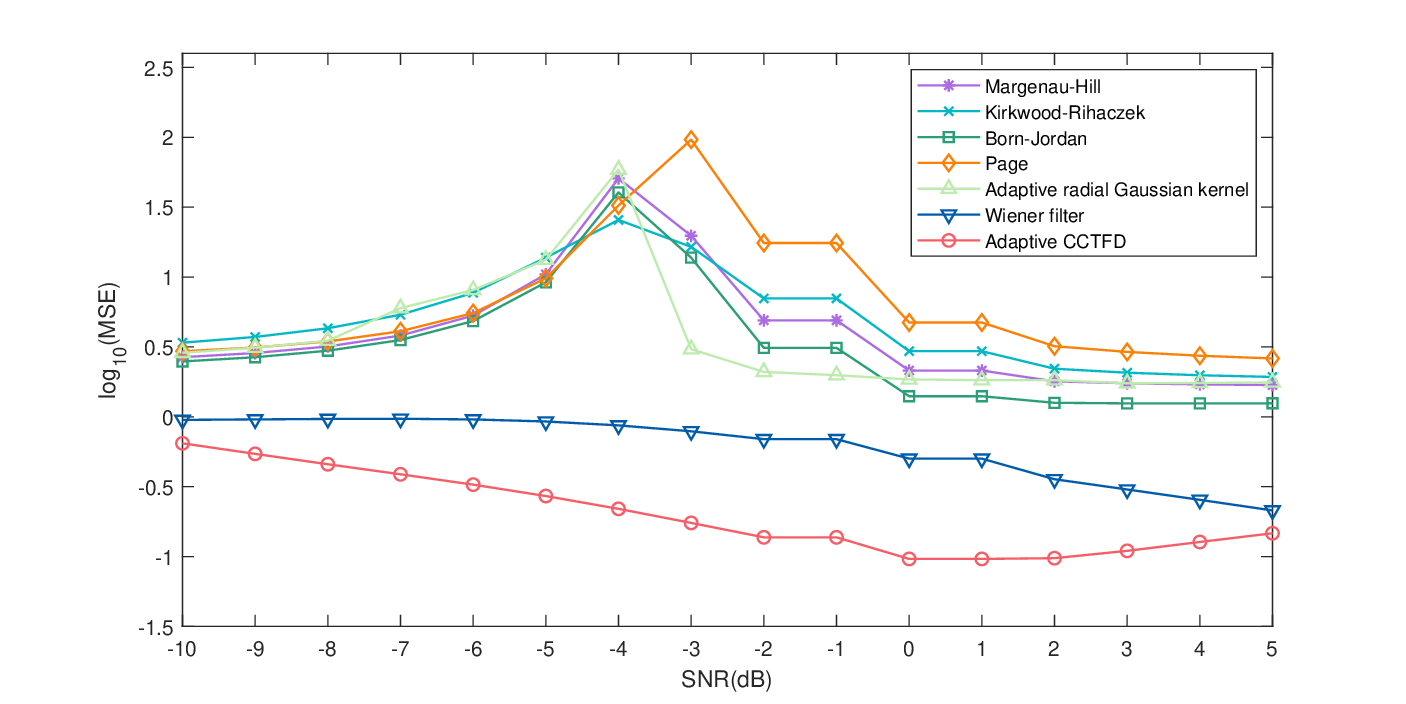}}
	\subfigure[]{\includegraphics[width=0.9\textwidth]{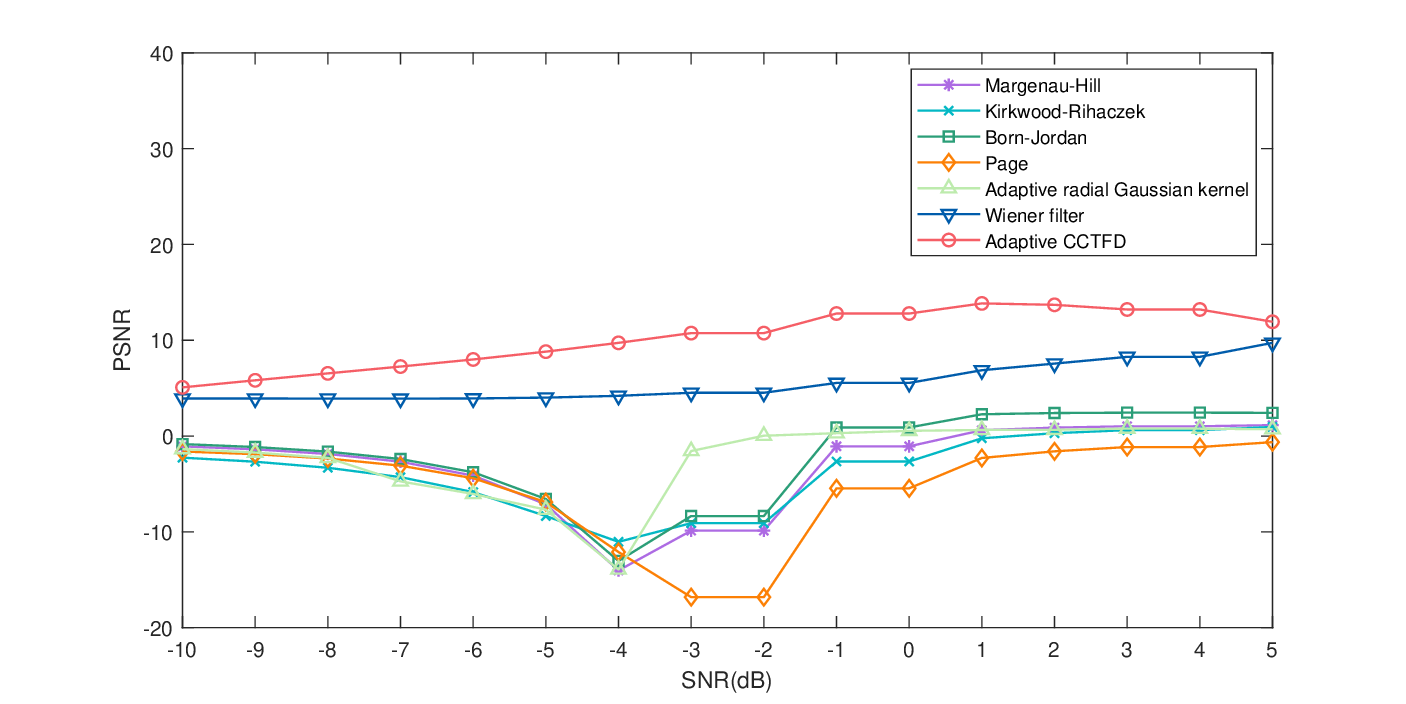}}
	\caption{Performance evaluation of seven denoising methods for the TC-LFM signal at SNR levels from $-10\mathrm{dB}$ to $5\mathrm{dB}$ under white Gaussian noise: (a) $\log_{10}(\text{MSE})$; (b) PSNR.}
	\label{fig4}
\end{figure}

\begin{figure}[!ht]
	\centering
	\includegraphics[width=1\linewidth]{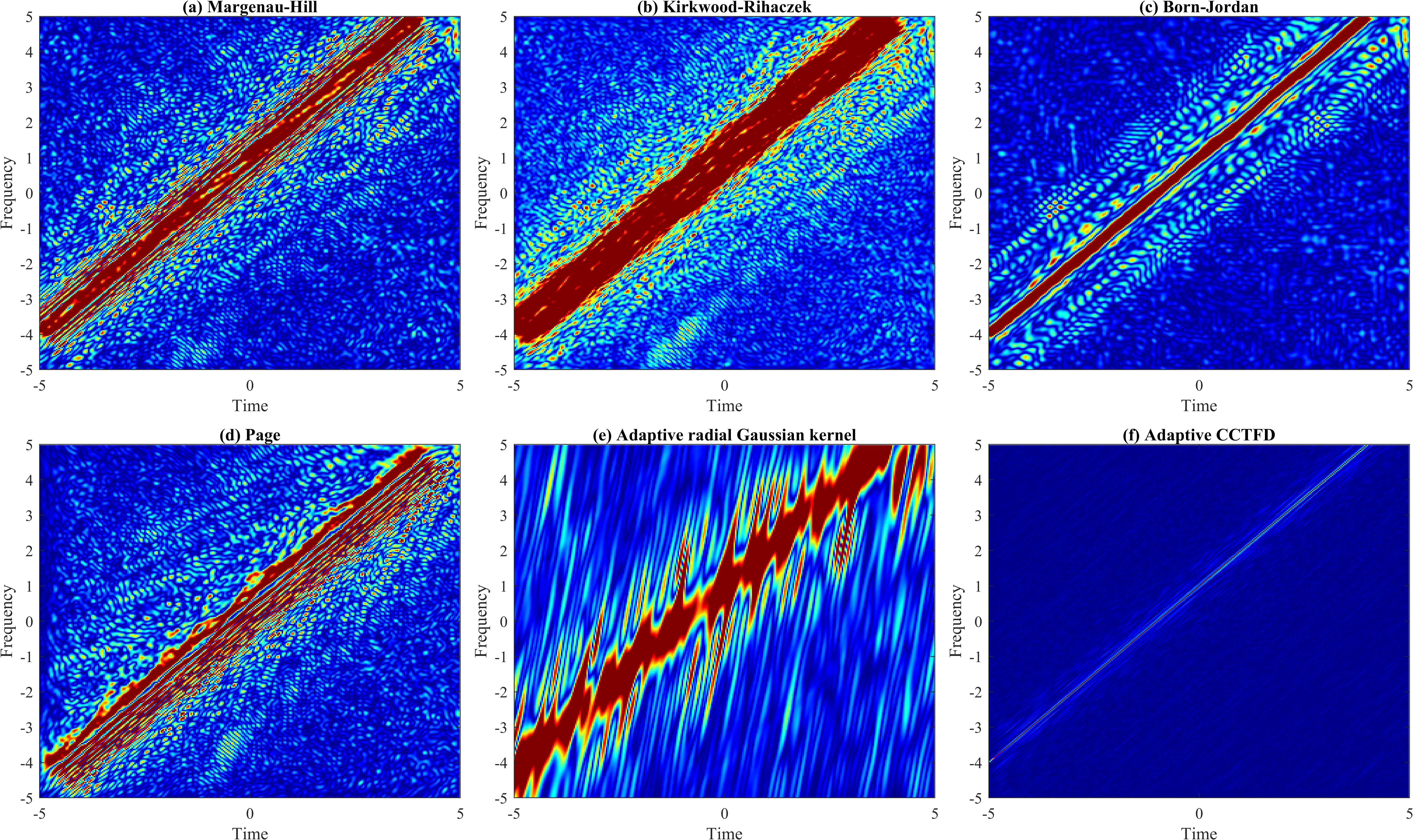}
	\caption{Visual representations of the TFDs for the estimated LFM signal at $\text{SNR} = 0 \mathrm{dB}$ using six different CCTFD methods.}
	\label{fig:TFD_comparison}
\end{figure}

{ To provide a more intuitive perspective on the filtering performance, Figure~\ref{fig:TFD_comparison} displays the visual representations of the 2D TFDs for the LFM signal at $\text{SNR} = 0\mathrm{dB}$. Note that the time-domain Wiener filter is excluded here as it does not generate a 2D TFD. As observed in Figures~\ref{fig:TFD_comparison}(a)--(d), the traditional fixed-kernel methods suffer from severe auto-term smearing, resulting in widened energy ridges. Furthermore, due to their rigid kernel designs, they fail to adequately suppress the background noise, leaving pervasive noise-induced artifacts and signal-noise cross-terms scattered across the time-frequency plane. The adaptive radial Gaussian kernel in Figure~\ref{fig:TFD_comparison}(e) still exhibits noticeable energy diffusion and strip-like artifacts. In sharp contrast, the proposed adaptive CCTFD shown in Figure~\ref{fig:TFD_comparison}(f) successfully isolates a sharp, concentrated energy trajectory with a clean background. This visual evidence perfectly corroborates the quantitative results, demonstrating the superior capability of the proposed method in preserving optimal time-frequency resolution while filtering out noise.}

{Furthermore, we investigate the denoising performance of the seven methods under different colored noise conditions, specifically pink, blue, and red noise. The observation intervals and sampling frequencies for the four signals remain identical to the previous settings.} Table~{\ref{color}} presents the MSE and PSNR values of the output signals through the seven filters under the influence of three types of colored noise. 
{Overall, the proposed adaptive CCTFD demonstrates robust superiority, achieving the best performance across all signals under blue noise, and for most signals (e.g., LFM and GELFM) under pink and red noise. The only exceptions occur under pink and red noise for signals with highly complex frequency dynamics (QFM and TC-LFM), where the Wiener filter exhibits an advantage. This performance divergence can be persuasively explained by examining the fundamental underlying assumptions and the distinct operational domains of the two methods. The Wiener filter operates under the assumption of 1D global wide-sense stationarity, minimizing the MSE based on the PSD. Because pink and red noises possess highly concentrated low-frequency PSDs, the Wiener filter can aggressively suppress these specific frequency bands without having to contend with time-frequency cross-term interference. Conversely, the proposed adaptive CCTFD is rigorously derived to achieve minimum MSE in the 2D WVD domain. Its underlying assumption is that the statistical properties—specifically, the 2D PSDs $\varepsilon_{W_f, W_g}$ and $\varepsilon_{W_g}$ in the ambiguity domain—can accurately characterize the optimal filter $H_{opt}(z)$. While this 2D statistical optimization is highly effective for signals like LFM and GELFM, it encounters unique theoretical challenges with highly non-linear (QFM) or multi-component crossing (TC-LFM) signals under strongly correlated colored noise. The WVDs of QFM and TC-LFM inherently contain severe, highly oscillatory cross-terms. When coupled with pink or red noise, the noise energy intensely interacts and couples with these signal cross-terms, creating highly convoluted 2D interference patterns. Consequently, the 2D statistical estimation required for synthesizing the optimal kernel $\phi_{opt}$ must contend with this complex cross-term interference, which slightly degrades its filtering precision. In contrast, the 1D Wiener filter completely bypasses 2D cross-term interference by operating solely on the 1D spectrum, granting it a slight macroscopic advantage in these specific, highly complex scenarios. Nonetheless, overall, our method consistently outperforms the fixed kernel filtering methods and the heuristic adaptive radial Gaussian kernel method.}

\begin{table}[htbp]
	
	\centering
	\caption{\label{color}The MSE and PSNR of seven filtering methods under different colored noise conditions.}
	\resizebox{0.65\textheight}{!}{
	\begin{tabular}{llllll}\hline
		~ & Example & LFM & GELFM & QFM & TC-LFM \\ \hline
		~ & \multicolumn{5}{c}{$\log_{10}(\text{MSE})$}  \\ \hline
		\multirow{7}*{Pink noise} & Margenau-Hill & $-0.0016$  & $-0.1592$  & $-0.0513$  & $0.2113$ \\ 
		~ & Kirkwood-Rihaczek & $0.0676$  & $-0.0050$  & $-0.0330$  & $0.2588$   \\  
		~ & Born-Jordan & $-0.1582$  & $-0.3053$  & $-0.1123$  & $0.0928$   \\ 
		~ & Page & $0.0996$  & $-0.0360$  & $-0.0201$  & $0.4003$   \\ 
		~ & Adaptive radial Gaussian kernel &  $-0.0451 $ &	$-0.4237$ &	$-0.0662$ &	$0.2232$  \\
		~ & Wiener filter & $-0.4985$  & $-0.6911 $ & $\mathbf{-0.4699}$  & $-0.4082$    \\ 
		~ & Adaptive CCTFD & $\mathbf{-1.5318}$  & $\mathbf{-0.9807}$  & $-0.3169$  & $\mathbf{-0.8599}$   \\   \hline
		\multirow{7}*{Blue noise} & Margenau-Hill & $0.1437$  & $-0.4563$  & $-0.0637$  & $0.2757$   \\ 
		~ & Kirkwood-Rihaczek & $0.3580$  & $-0.3711$  & $0.0063$  & $0.4053$   \\
		~ & Born-Jordan & $-0.1057$  & $-0.5265$  & $-0.2016$  & $0.0798$    \\ 
		~ & Page & $0.3281$  & $-0.3355$  & $0.0239$  & $0.5884$   \\ 
		~ & Adaptive radial Gaussian kernel &  $-0.0481 $ &	$-0.4892$ &	$-0.0671$ &	$0.3075$  \\
		~ & Wiener filter & $-0.4676$  & $-0.6632$  & $-0.5208$  & $-0.4113$   \\  
		~ & Adaptive CCTFD & $\mathbf{-1.2999}$  & $\mathbf{-0.9581}$  & $\mathbf{-0.7917}$  & $\mathbf{-1.0463}$   \\  \hline
		\multirow{7}*{Red noise} & Margenau-Hill & $0.0655$  & $-0.3004$  & $-0.0624$  & $0.2189$   \\ 
		~ & Kirkwood-Rihaczek & $0.1243$  & $-0.2567$  & $-0.0475$  & $0.2356$   \\ 
		~ & Born-Jordan & $-0.0907$  & $-0.3810$  & $-0.1369$  & $0.1392$  \\ 
		~ & Page & $0.1720$  & $-0.2790$  & $-0.0082$  & $0.3187$   \\ 
		~ & Adaptive radial Gaussian kernel &  $0.0105  $ &	$-0.3455$ &	$-0.0743$ &	$0.2559$  \\
		~ & Wiener filter & $-0.8039$  & $-1.2798$  & $\mathbf{-0.9489}$  & $\mathbf{-0.5090}$   \\ 
		~ & Adaptive CCTFD & $\mathbf{-1.7448}$  & $\mathbf{-1.3028}$  & $-0.4032$  & $-0.4568$ \\ \hline 
		~ & \multicolumn{5}{c}{PSNR} \\ \hline
		\multirow{7}*{Pink noise} & Margenau-Hill  & $3.0323$  & $5.2766$  & $3.5237$  & $1.2777$  \\ 
		~ & Kirkwood-Rihaczek & $2.3400$  & $3.7410$  & $3.3431$  & $1.2491$   \\  
		~ & Born-Jordan  & $4.6190$  & $7.0716$  & $4.1339$  & $2.4298$  \\ 
		~ & Page  & $2.0433$  & $3.7370$  & $3.2109$  & $-0.4254$ \\ 
		~ & Adaptive radial Gaussian kernel  & $3.4758   $ &	$7.3120 $ &	$3.6736 $ &	$0.9934$ \\
		~ & Wiener filter   & $8.8314$  & $10.3192$  & $\mathbf{8.0556}$  & $7.9690$   \\ 
		~ & Adaptive CCTFD   & $\mathbf{18.5390}$  & $\mathbf{12.8503}$  & $6.1978$  & $\mathbf{12.2002}$   \\   \hline
		\multirow{7}*{Blue noise} & Margenau-Hill  & $1.5865$  & $7.6228$  & $3.6468$  & $0.7775$  \\ 
		~ & Kirkwood-Rihaczek  & $-0.5454$  & $6.8108$  & $2.9469$  & $-0.1960$   \\
		~ & Born-Jordan  &  $4.1557$  & $8.3757$  & $5.0342$  & $2.5755$   \\ 
		~ & Page  & $-0.2493$  & $6.4764$  & $2.7708$  & $-2.2514$   \\ 
		~ & Adaptive radial Gaussian kernel & $3.5046 $ &	$7.9075 $ &	$3.6813  $ &	$-0.0112$ \\
		~ & Wiener filter  &  $8.1082$  & $9.8937$  & $8.6899$  & $9.1039$   \\  
		~ & Adaptive CCTFD & $\mathbf{16.0114}$  & $\mathbf{12.6000}$  & $\mathbf{10.9354}$  & $\mathbf{13.9650}$   \\  \hline
		\multirow{7}*{Red noise} & Margenau-Hill & $2.3888$  & $6.3998$  & $3.6360$  & $1.1878$   \\ 
		~ & Kirkwood-Rihaczek &  $1.7774$  & $5.9456$  & $3.4881$  & $1.3438$   \\ 
		~ & Born-Jordan &  $4.0111$  & $7.4167$  & $4.3805$  & $1.9829$   \\ 
		~ & Page & $1.3606$  & $5.9863$  & $3.0925$  & $0.3894$   \\ 
		~ & Adaptive radial Gaussian kernel & $2.9944 $ &	$6.6267  $ &	$3.7552 $ &	$0.8449$ \\
		~ & Wiener filter & $11.8832$  & $\mathbf{16.7052}$  & $\mathbf{13.1156}$  & $\mathbf{8.7673}$   \\ 
		~ & Adaptive CCTFD & $\mathbf{20.4654}$  & $16.1495$  & $7.0496$  & $8.0572$   \\ \hline 
	\end{tabular}
}
\end{table}

\indent {To further validate the practical effectiveness of the proposed adaptive CCTFD, we conduct experiments on real-world ECG data using the public ECG5000 dataset. Specifically, the $660$th sample of the dataset is extracted as the test signal and contaminated with additive white Gaussian noise under three specific SNR levels: $0\mathrm{dB}$, $1\mathrm{dB}$, and $2\mathrm{dB}$. For comparison, three representative neural network-based methods are also evaluated, including BernNet \cite{he2021bernnet}, LanczosNet \cite{liao2019lanczosnet}, and Specformer \cite{Bo23Specformer}. To ensure a fair comparison, these networks are trained on the preceding $659$ samples from the dataset. These training samples are augmented with corresponding noise levels and randomly split into $80\%$ for training and $20\%$ for validation. All baseline networks are trained for $100$ epochs with a learning rate of $0.001$. As shown in Table~\ref{ECG}, the proposed adaptive CCTFD consistently achieves the lowest MSE across all evaluated SNR levels. In terms of PSNR, while Specformer exhibits a marginal advantage specifically at $0\mathrm{dB}$, our method secures the highest PSNR under all other conditions. Overall, these results demonstrate that the adaptive CCTFD effectively preserves the crucial transient features  and the dynamic range of the ECG signals. }

\begin{table*}[!ht]
	
	\centering
	\caption{The MSE and PSNR of ten filtering methods for the ECG5000 dataset under white Gaussian noise.}
	\begin{tabular}{llccc}
		\toprule
		~ & Noise level (SNR) & $0$ dB & $1$ dB & $2$ dB \\ \midrule
		\multirow{15}*{$\log_{10}(\text{MSE})$} & \multicolumn{4}{c}{Time domain-based method} \\
		\cmidrule{2-5}
		~ & Wiener filter & $-0.6551$  & $-0.7027$  & $-0.7535$   \\ 
		\cmidrule{2-5}
		~ & \multicolumn{4}{c}{Time-frequency domain-based methods} \\
		\cmidrule{2-5}
		~ &  Margenau-Hill & $-0.3578$  & $-0.3986$  & $-0.4418$  \\
		~ & Kirkwood-Rihaczek & $-0.3296$  & $-0.3674$  & $-0.4073$   \\ 
		~ & Born-Jordan & $-0.5504$  & $-0.5867$  & $-0.6264$   \\ 
		~ & Page & $-0.2071$  & $-0.2573$  & $-0.3089$   \\ 
		~ & Adaptive radial Gaussian kernel & $-0.5765$  & $-0.6017$  & $-0.6122$   \\ 
		\cmidrule{2-5}
		~ & \multicolumn{4}{c}{Neural network-based methods} \\
		\cmidrule{2-5}
		~ & BernNet & $-0.4570$ & $-0.5374$ & $-0.6143$ \\
		~ & LanczosNet & $-0.3685$ & $-0.4156$ & $-0.4531$ \\
		~ & Specformer & $-0.4960$ & $-0.5195$ & $-0.5393$ \\
		\cmidrule{2-5}
		~ & Adaptive CCTFD & $\mathbf{-0.6987}$  & $\mathbf{-0.7682}$  & $\mathbf{-0.8328}$   \\ \bottomrule 
		\toprule
		\multirow{15}*{PSNR} & \multicolumn{4}{c}{Time domain-based method} \\
		\cmidrule{2-5}
		~ & Wiener filter & $7.6116$  & $7.8973$  & $8.1409$   \\ 
		\cmidrule{2-5}
		~ & \multicolumn{4}{c}{Time-frequency domain-based methods} \\
		\cmidrule{2-5}
		~ & Margenau-Hill & $5.2520$  & $5.2621$  & $5.3176$   \\ 
		~ & Kirkwood-Rihaczek & $5.6477$  & $5.6275$  & $5.6327$   \\ 
		~ & Born-Jordan & $6.0279$  & $6.1368$  & $6.4350$   \\ 
		~ & Page & $5.0387$  & $4.8682$  & $4.8384$   \\ 
		~ & Adaptive radial Gaussian kernel & $6.1222$  & $6.4040$  & $6.5147$   \\ 
		\cmidrule{2-5}
		~ & \multicolumn{4}{c}{Neural network-based methods} \\
		\cmidrule{2-5}
		~ & BernNet & $7.1874$ & $7.6312$ & $8.2194$ \\
		~ & LanczosNet & $5.4695$ & $5.9339$ & $6.3003$ \\
		~ & Specformer & $\mathbf{8.3500}$ & $8.2515$ & $8.1368$ \\
		\cmidrule{2-5}
		~ & Adaptive CCTFD & $7.7642$  & $\mathbf{8.4248}$  & $\mathbf{9.0422}$   \\ \bottomrule 
		\label{ECG}
	\end{tabular}
\end{table*}

\paragraph{ Advantages and limitations}
The proposed adaptive CCTFD framework offers several notable advantages compared with existing time-frequency analysis methods. First, by integrating Wiener filtering theory within the CCTFD structure, the proposed method achieves statistical optimality under the MMSE criterion, which is particularly effective for signal denoising in low {signal-to-noise} ratio conditions. Second, unlike {decomposition-based} approaches such as FBSE, tunable-Q wavelet, or Hankel matrix based techniques, which often involve iterative procedures or basis selection, the proposed approach follows a {non-iterative} filtering paradigm in the time-frequency domain, leading to a conceptually simpler and more direct denoising mechanism. Additionally, the convolutional form of CCTFD preserves interpretability of the time-frequency representation while enabling adaptive kernel design tailored to signal characteristics.

Despite these advantages, the proposed method has limitations. The derivation of the adaptive kernel relies on {second-order} statistical assumptions inherent to Wiener filtering, which may limit performance for signals with strongly {non-Gaussian higher-order} statistics. {Regarding the execution time, because the proposed method operates on a 2D time-frequency plane, it naturally sacrifices some speed compared to 1D time-domain techniques like the Wiener filter. However, among 2D time-frequency methods, it exhibits significant computational efficiency and scalability. To rigorously evaluate this, a Monte Carlo test is conducted by averaging the execution times over 20 independent trials for random signals of varying lengths ($T = 250, 500, 750, 1000$).As summarized in Table \ref{tab:running_time}, the proposed adaptive CCTFD significantly reduces the execution time compared to the iterative adaptive radial Gaussian kernel (e.g., reducing the time by more than 60\% at $N=1000$). Furthermore, it even requires slightly less computational time than the traditional fixed-kernel methods across all tested signal lengths. This efficiency indicates that deriving the adaptive kernel directly in the WVD domain under the least-squares criterion avoids the heavy computational burden typical of conventional adaptive kernel designs.
}

\begin{table}[htbp]
	
	\centering
	\caption{Average execution times (in seconds) for various signal lengths $T$.}
	\label{tab:running_time}
	\begin{tabular}{lcccc}
		\hline
		Method & $T=250$ & $T=500$ & $T=750$ & $T=1000$ \\
		\hline
		Margenau-Hill                   & $1.4289$ & $5.3032$  & $15.3119$ & $21.9656$ \\
		Kirkwood-Rihaczek               & $1.5578$ & $5.5453$  & $15.7876$ & $22.2232$ \\
		Born-Jordan                     & $1.4552$ & $5.3277$  & $15.7318$ & $22.0931$ \\
		Page                            & $1.5700$ & $5.5524$  & $16.3405$ & $22.2814$ \\
		Adaptive radial Gaussian kernel & $2.2525$ & $12.1010$ & $37.5738$ & $61.6663$ \\
		Wiener filter                   & $0.0009$ & $0.0009$  & $0.0010$  & $0.0008$  \\
		Adaptive CCTFD                  & $0.7538$ & $3.6061$  & $12.3473$ & $18.9453$ \\
		\hline
	\end{tabular}
\end{table}

\section{Conclusion}\label{sec:5}
\indent {A convolutional CCTFD time-frequency analysis framework, explicitly tailored for the adaptive denoising of signals corrupted by additive noise under low SNR conditions, was established. This approach yields an adaptive kernel function that minimizes the MSE in the WVD domain.} The proposed adaptive CCTFD can automatically adjust its kernel function according to the change of signal to adapt to different signal characteristics. 
{Experimental results demonstrate that its denoising performance} is superior not only to some classical fixed kernel function-based CCTFDs, including the Margenau-Hill distribution, the Kirkwood-Rihaczek distribution, the Born-Jordan distribution, and the Page distribution, but also to the adaptive radial Gaussian kernel function-based CCTFD. Under white Gaussian noise, our method outperforms the ordinary Wiener filter, while under colored noise, it performs comparably to the Wiener filter. {Experiments on real-world {ECG} data further demonstrate that the proposed adaptive CCTFD consistently outperforms both traditional signal processing methods and representative neural network-based approaches in terms of MSE and PSNR, confirming its practical applicability and robustness.}

{
	\section{Future work}\label{sec:6}
	\indent Recently, there has been increasing interest in multichannel and multivariate time-frequency analysis for complex signal processing tasks, such as multivariate empirical Fourier decomposition, empirical wavelet transform, and iterative filtering. The present work focuses on the single-channel case, where the adaptive kernel of Cohen's class time-frequency distribution is designed in the WVD domain under the minimum mean-square error criterion.
	
	An important direction for future research is the extension of the proposed framework to multichannel signal analysis. From a theoretical perspective, the proposed least-squares adaptive filtering strategy can be naturally generalized by introducing multichannel or cross WVD and matrix-valued power spectral density functions. In this case, the adaptive kernel function becomes matrix-valued and can be optimized via a vector Wiener filtering principle to jointly exploit inter-channel correlations in the time-frequency domain.
	
	Such a multichannel adaptive CCTFD framework would enable joint denoising and representation of multivariate nonstationary signals, providing a complementary approach to existing multivariate decomposition-based methods such as multivariate EFD, EWT, and iterative filtering. Exploring efficient implementations, robustness to channel-dependent noise, and comparisons with these state-of-the-art multichannel techniques will be the subject of future work.
}


\appendix
\section{Proof of Eq.~\eqref{eq:10}}
\indent By using the orthogonal principle
\begin{align}\label{eq:A.1}
	\mathbb{E}\left\{\left[\mathrm{W}_f(\boldsymbol{z})-\left(\mathrm{W}_g\ast H_{\mathrm{opt}}\right)(\boldsymbol{z})\right]\overline{\mathrm{W}_g(\boldsymbol{z'})}\right\}=0, 
	\boldsymbol{z'}\in\mathbb{R}^{2N},
	\tag{A.1}
\end{align}
we establish the Wiener-Hopf equation
\begin{equation}\label{eq:A.2}
	R_{\mathrm{W}_f,\mathrm{W}_g}(\boldsymbol{z},\boldsymbol{z'})-\int_{\mathbb{R}^{2N}}R_{\mathrm{W}_g}(\boldsymbol{k},\boldsymbol{z'})H_{\mathrm{opt}}(\boldsymbol{z}-\boldsymbol{k})\mathrm{d}\boldsymbol{k}=0,
	\tag{A.2}
\end{equation}
where $R_{\mathrm{W}_f,\mathrm{W}_g}$ denotes the cross-correlation function between $\mathrm{W}_f$ and $\mathrm{W}_g$, and $R_{\mathrm{W}_g}$ denotes the auto-correlation function of $\mathrm{W}_g$. In general, we can obtain $H_{\mathrm{opt}}$ by solving Eq.~\eqref{eq:A.2} numerically. Particularly, if $\mathrm{W}_f$ and $\mathrm{W}_g$ are stationary, Eq.~\eqref{eq:A.2} simplifies to
\begin{equation}\label{eq:A.3}
	R_{\mathrm{W}_f,\mathrm{W}_g}(\boldsymbol{z}-\boldsymbol{z'})-\int_{\mathbb{R}^{2N}}R_{\mathrm{W}_g}(\boldsymbol{k}-\boldsymbol{z'})H_{\mathrm{opt}}(\boldsymbol{z}-\boldsymbol{k})\mathrm{d}\boldsymbol{k}=0.
	\tag{A.3}
\end{equation}
\indent Taking the change of variables $\boldsymbol{z}-\boldsymbol{z'}=\boldsymbol{p}$ and $\boldsymbol{z}-\boldsymbol{k}=\boldsymbol{q}$ yields
\begin{equation}\label{eq:A.4}
	R_{\mathrm{W}_f,\mathrm{W}_g}(\boldsymbol{p})-\int_{\mathbb{R}^{2N}}R_{\mathrm{W}_g}(\boldsymbol{p}-\boldsymbol{q})H_{\mathrm{opt}}(\boldsymbol{q})\mathrm{d}\boldsymbol{q}=0.
	\tag{A.4}
\end{equation}
\indent Thanks to the conventional convolution and correlation theorems, we solve Eq.~\eqref{eq:A.4} to obtain
\begin{equation}\label{eq:A.5}
	\mathcal{F}\left[\mathrm{W}_f\right](\boldsymbol{u})\overline{\mathcal{F}\left[\mathrm{W}_g\right](\boldsymbol{u})}=\mathcal{F}\left[H_{\mathrm{opt}}\right](\boldsymbol{u})\left|\mathcal{F}\left[\mathrm{W}_g\right](\boldsymbol{u})\right|^2,
	\tag{A.5}
\end{equation}
and therefore, we arrive the required result \eqref{eq:10}.$\hfill\blacksquare$

\section{Proof of Eq.~\eqref{eq:12}}  
\indent The minimize MSE takes
\begin{equation}\label{eq:B.1}
	\mathop{\min}\limits_{H(\boldsymbol{z})}\sigma_{\mathrm{MSE}}^{2}=\mathbb{E}\left\{\left[\mathrm{W}_f(\boldsymbol{z})-\left(\mathrm{W}_g\ast H_{\mathrm{opt}}\right)(\boldsymbol{z})\right]\overline{\mathrm{W}_f(\boldsymbol{z})}\right\}.
	\tag{B.1}
\end{equation}
\indent Similar to Eqs.~\eqref{eq:A.2}--\eqref{eq:A.4}, we have
\begin{equation}\label{eq:B.2}
	\mathop{\min}\limits_{H(\boldsymbol{z})}\sigma_{\mathrm{MSE}}^{2}=R_{\mathrm{W}_f}(\boldsymbol{0})-\int_{\mathbb{R}^{2N}}R_{\mathrm{W}_g,\mathrm{W}_f}(-\boldsymbol{k})H_{\mathrm{opt}}(\boldsymbol{k})\mathrm{d}\boldsymbol{k}.
	\tag{B.2}
\end{equation}
\indent Because of Parseval's relation of the WVD, it follows that
\begin{equation}\label{eq:B.3}
	R_{\mathrm{W}_f}(\boldsymbol{0})=\int_{\mathbb{R}^{2N}}\left|\mathrm{W}_f(\boldsymbol{z})\right|^2\mathrm{d}\boldsymbol{z}=\left\|f\right\|_2^4.
	\tag{B.3}
\end{equation}
\indent Thanks to the conventional convolution and correlation theorems, substituting Eq.~\eqref{eq:10} yields

\begin{align}\label{eq:B.4}
	\int_{\mathbb{R}^{2N}}R_{\mathrm{W}_g,\mathrm{W}_f}(-\boldsymbol{k})H_{\mathrm{opt}}(\boldsymbol{k})\mathrm{d}\boldsymbol{k} 
	=&\int_{\mathbb{R}^{2N}}\mathcal{F}\left[H_{\mathrm{opt}}\right](\boldsymbol{u})\mathcal{F}\left[\mathrm{W}_g\right](\boldsymbol{u})\overline{\mathcal{F}\left[\mathrm{W}_f\right](\boldsymbol{u})}\mathrm{d}\boldsymbol{u} \nonumber\\
	=&\int_{\mathbb{R}^{2N}}\frac{\varepsilon_{\mathrm{W}_f,\mathrm{W}_g}(\boldsymbol{u})\varepsilon_{\mathrm{W}_g,\mathrm{W}_f}(\boldsymbol{u})}{\varepsilon_{\mathrm{W}_g}(\boldsymbol{u})}\mathrm{d}\boldsymbol{u}.
	\tag{B.4}
\end{align}
\indent With Eqs.~\eqref{eq:B.2}, \eqref{eq:B.3} and \eqref{eq:B.4}, we have
\begin{equation}
	\mathop{\min}\limits_{H(\boldsymbol{z})}\sigma_{\mathrm{MSE}}^{2}=\left\|f\right\|_2^4-\int_{\mathbb{R}^{2N}}\frac{\varepsilon_{\mathrm{W}_f,\mathrm{W}_g}(\boldsymbol{u})\varepsilon_{\mathrm{W}_g,\mathrm{W}_f}(\boldsymbol{u})}{\varepsilon_{\mathrm{W}_g}(\boldsymbol{u})}\mathrm{d}\boldsymbol{u}.
	\tag{B.5}
\end{equation}
\indent By further simplifying the above equation, we arrive the required result \eqref{eq:12}.  $\hfill\blacksquare$




\section*{Reference} 
  \bibliographystyle{elsarticle-num-names}
  \bibliography{reference}





\end{sloppypar}
\end{document}